\documentclass{jpsj3}

\title{Anomalous Coexistence of Ferroelectric Phases ($P\parallel a$ and $P\parallel c$) in Orthorhombic Eu$_{1-y}$Y$_y$MnO$_3$ ($y>0.5$) Crystals}

\author{Mitsuru \textsc{Akaki}\thanks{E-mail: m-akaki@sophia.ac.jp}, Masaaki \textsc{Hitomi}, Mizuaki \textsc{Ehara}, $^{1}$Daisuke \textsc{Akahoshi}, and Hideki \textsc{Kuwahara}}

\inst{Department of Physics, Sophia University, Tokyo 102-8554, Japan \\
$^{1}$ Department of Physics, Toho University, Funabashi 274-8510, Japan}

\abst{We have investigated the magnetic and dielectric properties of orthorhombic Eu$_{1-y}$Y$_y$MnO$_3$ ($0\leq y\leq 0.6$) single crystals without the
presence of the 4$f$ magnetic moments of the rare-earth ions. In $y\geq 0.2$, the magnetic-structure driven ferroelectricity is observed. The ferroelectric transition temperature is steeply reducing with increasing $y$. In $y\geq 0.52$, two ferroelectric phases ($P\parallel a$ and $P\parallel c$) are coexistent at low temperatures. In these phases, ferroelectricity has different origin, which is evidenced by the distinctive poling-electric-field dependence of electric polarization. Namely, the electric polarization along the $c$ axis ($P_c$) is easily saturated by a poling electric field, therefore $P_c$ is caused by the $bc$ spiral antiferromagnetic order. On the other hand, the electric polarization along the $a$ axis ($P_a$) is probably attributed to the collinear $E$-type antiferromagnetic order, because $P_a$ is unsaturated even in a poling field of $10^6$ V/m.
}

\kword{multiferroic, magnetoelectric property, perovskite manganese oxide, phase diagram}

\begin{document}
\maketitle

\newpage
\section{Introduction}
The observation of ferroelectricity accompanied by a magnetic order in TbMnO$_3$ has led to a renewed interest in research on magnetoelectric multiferroics.\cite{nature}
$R$MnO$_3$ ($R$: rare earth) with an orthorhombic perovskite structure is one of the most typical examples of the recently-discovered multiferroic materials. 
In $R$~=~La$-$Eu, $R$MnO$_3$ has a collinear $A$-type antiferromagnetic (AFM) ground state of Mn 3$d$ spins. With decreasing the $R$-site ionic radius, the magnetic ground state changes from the $A$-type AFM order to a noncollinear transverse spiral AFM one around $R$~=~Gd, owing to the magnetic frustration between the nearest neighbor (NN) ferromagnetic and the next NN AFM interactions among Mn spins.\cite{RMnO3_Kim,RMnO3_Got}
In $R$MnO$_3$ ($R$~=~Tb,~Dy), the noncollinear transverse spiral AFM order breaks the inversion symmetry through the inverse Dzyaloshinskii$-$Moriya (DM) interaction, resulting in the appearance of ferroelectricity.\cite{spiral_Kat,spiral_Mos,spiral_Ken,spiral_Ari}
In other words, in the transverse spiral spin structure, the spin chirality can induce the electric polarization 
$\mbox{\boldmath$P$}\propto \textstyle\sum\limits_{ij} \mbox{\boldmath $e$}_{ij}\times (\mbox{\boldmath $S$}_i\times \mbox{\boldmath $S$}_j)$ in terms of the spin current mechanism, where $\mbox{\boldmath $e$}_{ij}$ denotes the unit vector connecting the interacting neighbor spins $\mbox{\boldmath$S$}_i$ and $\mbox{\boldmath$S$}_j$.
The spin helicity or vector spin chirality of the spiral spin structure can be changed by external magnetic fields, and then the direction of the ferroelectric polarization due to the spin chirality can also be controlled by the fields.\cite{s_chirality}
With further decreasing the $R$-site ionic radius from Dy, the ground state changes from the noncollinear spiral AFM order to a collinear $E$-type AFM one ($R$~=~Ho$-$Lu,~and~Y). $R$MnO$_3$ with the $E$-type AFM order also exhibits ferroelectric behavior, which is caused not by the inverse DM interaction but by the inverse Goodenough$-$Kanamori (GK) interaction, that is, the symmetric exchange striction term expressed as $\mbox{\boldmath$S$}_i \cdot \mbox{\boldmath$S$}_j$.\cite{Etype_theo,Etype_poly,Etype_Ishi}
This type of ferroelectricity is also realized in DyFeO$_3$\cite{DyFeO3} and Ca$_3$(Co,Mn)$_2$O$_6$\cite{CaCoMnO}.

The magnetoelectricity of $R$MnO$_3$ is largely affected not only by the magnetic frustration of Mn 3$d$ spins but also by 4$f$ magnetic moments of $R^{3+}$.
A part of the results on Eu$_{1-y}$Y$_y$MnO$_3$ that excludes the influence of 4$f$ moments by using Eu$^{3+}$ ($J=0$) and Y$^{3+}$ (nonmagnetic) has been reported in our preceding paper.\cite{EuY_Noda} Furthermore, details of the effect of 4$f$ moments were clarified from the comparative study of (Eu,Y)$_{1-y}$Tb$_y$MnO$_3$ and (Eu,Y)$_{1-y}$Gd$_y$MnO$_3$ crystals with different 4$f$ magnetic characters.\cite{Hitomi}
The results of our experiments clearly show that the ferroelectricity of $R$MnO$_3$ is greatly influenced by the existence of a small amount of 4$f$ moment. 
Therefore, in this study, we focused on 4$f$-moment-free system Eu$_{1-y}$Y$_y$MnO$_3$.
The ground state of Eu$_{1-y}$Y$_y$MnO$_3$ changes from a canted $A$-type AFM and paraelectric phase with concentration $y < 0.15$ toward a presumably spiral AFM and ferroelectric one with $y \ge 0.3$ (see also Fig.~\ref{fig4}).\cite{EuY_Hem,EuY_Yam}
In the case of $y=0.405$ for Eu$_{1-y}$Y$_y$MnO$_3$, the direction of the ferroelectric polarization spontaneously changes from along the $c$-axis ($P_c$) at higher temperatures to the $a$-axis ($P_a$) at lower temperatures in a zero magnetic field.\cite{EuY_Noda}
In the ferroelectric phase of $y=0.405$, the spiral spin structure is confirmed by neutron scattering measurements.\cite{EuY_Kaji}
 However, $P_c$ is a minor phase and its magnitude is much smaller than that of $P_a$ in the $y=0.405$ sample.
To clarify the phase competition between the $P_a$ and the $P_c$ states, it is necessary to develop the $P_c$ phase and to enhance its magnitude by increasing $y$. This is become the $P_c$ region expands with increasing Y concentration $y$ as shown in Fig.~\ref{fig4}.
In addition, with increasing $y$ or equivalently decreasing the average ionic radius of $R$ site, the emergence of the ferroelectric state with the $E$-type AFM order is believed to occur, as observed in polycrystalline form.\cite{Etype_Ishi}
In this study, to reveal the electronic phase separation or coexistence between the $P_a$ and the $P_c$ phases as well as to achieve the ferroelectric phase with the $E$-type AFM order, we have investigated the magnetic and ferroelectric properties of orthorhombic Eu$_{1-y}$Y$_y$MnO$_3$ (especially~for~$y > 0.5$) single crystals, in which development of the $P_c$ phase and the emergence of $E$-type AFM order are expected.

\section{Experiment}
We prepared a series of Eu$_{1-y}$Y$_y$MnO$_3$ ($0\leq y\leq 0.6$) single crystals which were grown by a floating zone method.
In order to stabilize the orthorhombic phase, we used high-pressure oxygen of 7.5 atm as a growth atmosphere for $0.5\leq y\leq 0.6$ crystals, while argon of 2.5 atm for $0\leq y<0.5$ crystals.
We performed X-ray diffraction (XRD) measurements on the obtained crystals at room temperature. The Rietveld analysis of powder XRD data revealed that all the samples have an orthorhombic $Pbnm$ structure without a hexagonal impurity phase or any phase segregation. All the single crystalline samples used in this study were cut along the crystallographic principal axes into a platelike shape using an X-ray back-reflection Laue technique. We also prepared the sample oriented along the [101] direction to investigate the phase coexistence of $P_a$ and $P_c$. 
The dielectric constant and the ferroelectric polarization were measured using a temperature-controllable cryostat. 
The dielectric constant measurement was performed with an $LCR$ meter at a frequency of 10~kHz (Agilent,~4284A). The pyroelectric current to obtain the spontaneous electric polarization was measured in a warming process at a rate of 4~K/min after the samples were cooled from 60~to~5~K in a poling electric field of 20~$\sim$~1020~kV/m. The magnetic properties and specific heat were measured using a commercial apparatus (Quantum~Design,~PPMS-9T).

\section{Results and discussion}
\begin{figure}
\begin{center}
\includegraphics[scale=0.60, clip]{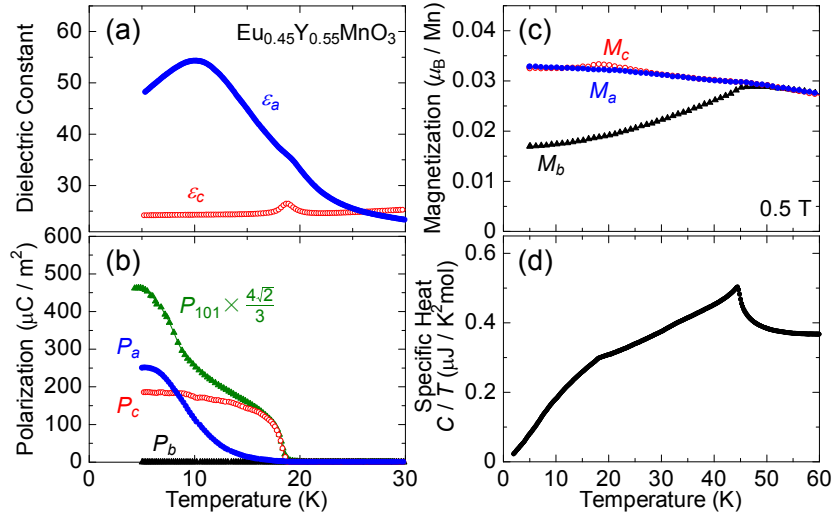}
\vspace{-9pt}
\caption{\label{fig1} (Color online) Temperature dependence of (a) dielectric constant along the $a$- and $c$-axes, (b) electric polarization along the $a$, $b$, $c$, and [101] directions, (c) magnetization along the each principal axis, and (d) specific heat divided by temperature of Eu$_{0.45}$Y$_{0.55}$MnO$_3$. To compare the $(P_a+P_c)$ value, $P_{101}$ is multiplied by a factor of $\frac{4\sqrt{2}}{3}$.\cite{P101} The magnetization was measured in 0.5~T\@.}
\end{center}
\end{figure}

Let us start with the results of Eu$_{0.45}$Y$_{0.55}$MnO$_3$. Figure \ref{fig1} shows the temperature dependence of (a) dielectric constant, (b) electric polarization, (c) magnetization, and (d) specific heat divided by temperature $C/T$. 
The sample shows the anisotropic magnetization below 45~K, above which a common paramagnetic behavior is observed along all axes. Below 45~K, the magnetization parallel to the $b$-axis ($M_b$) steeply decreases whereas that along the $a$ ($M_a$)- and $c$ ($M_c$)- axes remains almost constant in the temperature range from 45 down to 18~K\@. This result implies that the AFM order has a magnetic easy axis parallel to the $b$-axis.
$P_c$ rises up sharply below 18~K, where the dielectric constant along the $c$-axis ($\varepsilon _c$) shows a small peak concomitantly. The slight anomaly is observed also in the $M_c$ and $C/T$ around the same ferroelectric transition temperature. On the other hand, the $P_a$ phase develops gradually below 18~K, where $\varepsilon _a$ shows a faint anomaly. 
A broad peak of $\varepsilon _a$ is observed around 10~K, below which coexistence of the $P_a$ and the $P_c$ phases is prominent.
To ensure that this coexistence is not an experimental artifact such as misalignment of the sample direction, we have measured the ferroelectric polarization along the [101] direction (Fig. \ref{fig1} (b)).
Then, we have observed the distinctive feature of $P_{101}$, which is consistent with the expected ($P_a+P_c$) value projected on to the [101] direction.
This is in sharp contrast to the case of the $y=0.405$ sample, in which the emergence of the $P_a$ phase and the disappearance of the $P_c$ phase occurred simultaneously at a certain temperature.\cite{EuY_Noda}
In other words, the temperature-induced electric polarization flop, i.e., 90$^{\circ}$-rotation of the polarization, was observed.
Therefore, there is no indication of coexistence of both the $P_a$ and the $P_c$ phases in the $y=0.405$ sample.

\begin{figure}
\begin{center}
\includegraphics[scale=0.41, clip]{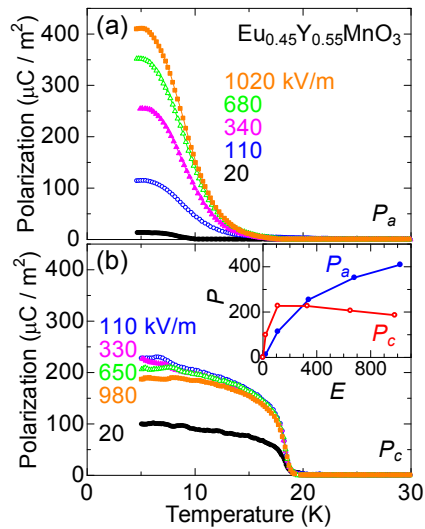}
\vspace{-9pt}
\caption{\label{fig2} (Color online) Temperature dependence of ferroelectric polarization along (a) the $a$-axis and (b) the $c$-axis of Eu$_{0.45}$Y$_{0.55}$MnO$_3$ crystal in various poling electric fields, which were applied in the cooling process and removed before the measurements of polarization. The inset shows the value of ferroelectric polarization along the $a$- and $c$-axes at 5 K as a function of the poling electric field.}
\end{center}
\end{figure}

Figure~\ref{fig2} shows the temperature dependence of $P_a$ and $P_c$ of Eu$_{0.45}$Y$_{0.55}$MnO$_3$ in various poling electric fields. 
$P_c$ is saturated at a field of 110~kV/m with a value of about 230~$\mu$C/m$^{2}$. 
On the other hand, $P_a$ is not saturated even in a field of 1020~kV/m. $P_a$ can possibly indicate a larger magnitude in a stronger poling field.
The observed large difference in saturation electric fields of the $P_a$ and the $P_c$ phases suggests that each phase has a different origin of its ferroelectricity. 
Namely, the $P_c$ phase originates from the inverse DM interaction for the $bc$-spiral spin structure. 
A saturation poling field of about 100~kV/m for the $P_c$ phase is comparable to those for other ferroelectric phases due to the inverse DM interaction. 
On the other hand, the $P_a$ phase seems to arise from the inverse GK interaction for the $E$-type AFM order which is expected to exist in such high-$y$ samples.\cite{EuY_theo} 
The similar poling field dependence of $P_a$, that is, no saturation of $P_a$ up to 500~kV/m, is reported in a polycrystalline YMnO$_3$ sample in which the $E$-type AFM order is confirmed by neutron scattering.\cite{Etype_poly} 

Therefore, the data shown in Figs.~\ref{fig1}~and~\ref{fig2} clearly evidence that the $P_a$ and the $P_c$ phases with different origins coexist in low temperatures especially below 10~K\@.
These results indicate that this coexisting state originates from the difference of the AFM orders that cause $P_a$ or $P_c$. In the $y=0.60$ sample, a similar poling field dependence to that of $y=0.55$ was observed.

\begin{figure}
\begin{center}
\includegraphics[scale=0.41, clip]{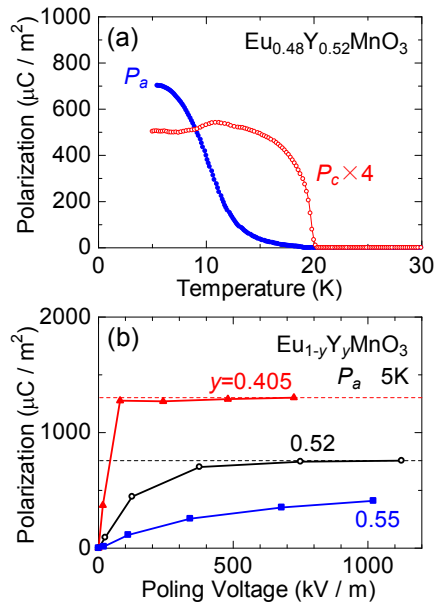}
\vspace{-9pt}
\caption{\label{fig3} (Color online) (a) Temperature dependence of ferroelectric polarization along the $a$- and $c$-axes of $y=0.52$. $P_c$ is multiplied by a factor of 4 for comparison. (b) The value of ferroelectric polarization along the $a$-axis of $y=0.405$, 0.52, and 0.55 at 5 K as a function of the poling electric field. The dotted lines are guides to the eye.}
\end{center}
\end{figure}

\begin{figure}
\begin{center}
\includegraphics[scale=0.4, clip]{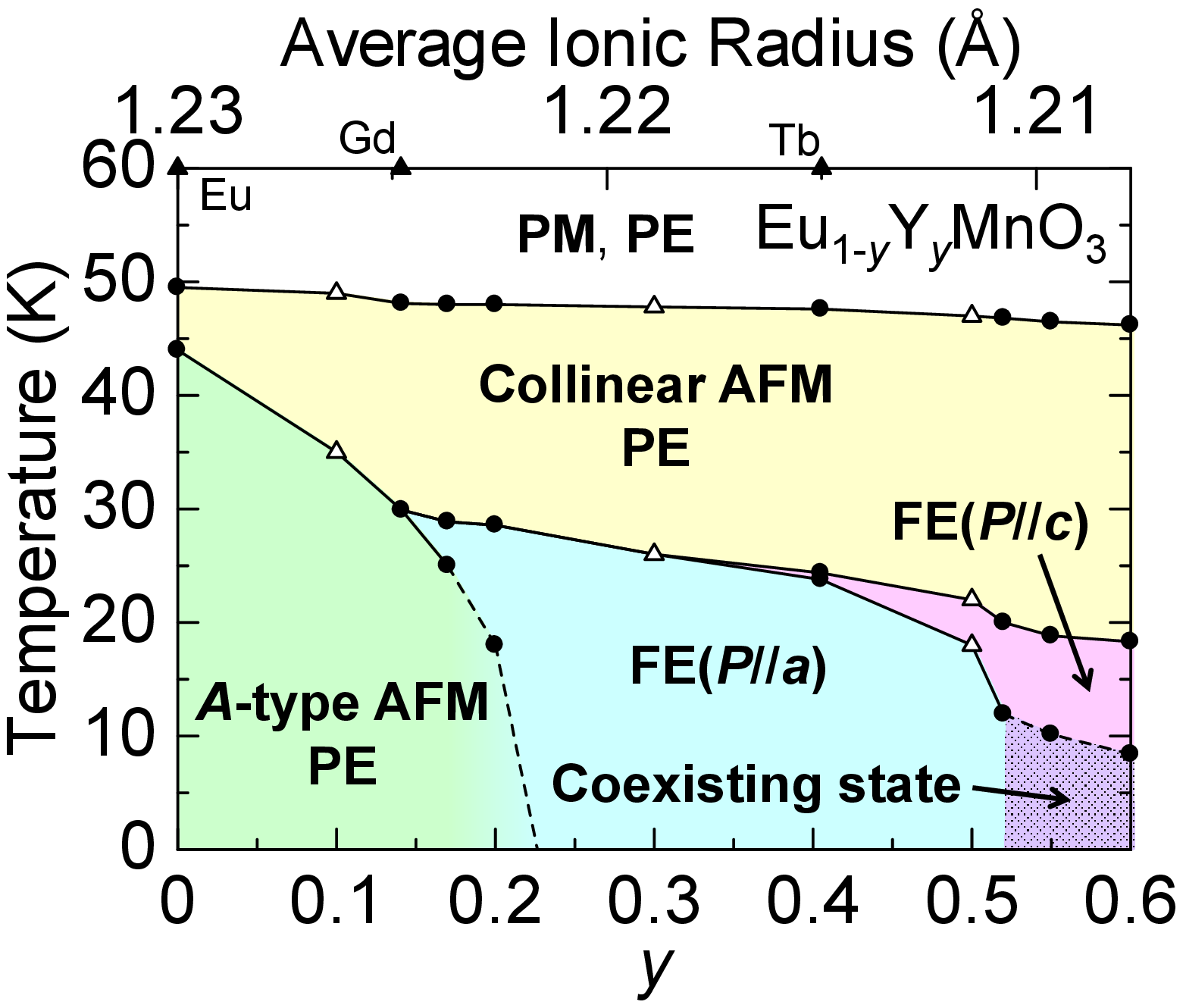}
\vspace{-9pt}
\caption{\label{fig4} (Color online) The magnetoelectric phase diagram of Eu$_{1-y}$Y$_y$MnO$_3$ with 0~$\le~y~\le$~0.6\@. The abbreviations mean paramagnetic (PM), paraelectric (PE), antiferromagnetic (AFM), and ferroelectric (FE) phases. The data points are obtained from the measurements of the dielectric constant, and from Ref.~16~($\bigtriangleup $). The hatched area in the right bottom region means the phase-coexisting state. The upper horizontal axis indicates the average ionic radius of (Eu$_{1-y}$Y$_y$) ion, which corresponds to the $y$ value in lower axis. Solid triangles in upper axis denote the ionic radius of each rare earth ion.}
\end{center}
\end{figure}

Figure~\ref{fig3} (a) shows the temperature dependence of $P_a$ and $P_c$ of $y=0.52$ located near the phase boundary. The $P_a$ and the $P_c$ phases of $y=0.52$ show a similar behavior to $y=0.55$ (see also Fig.~\ref{fig1} (b)); therefore, we conclude that the $P_a$ phase of $y=0.52$ is also induced by the $E$-type AFM order. However, in contrast to the $y=0.55$, $P_c$ is strongly sappressed and slightly decreases below 10~K where $P_a$ sharply rises up. This polarization-flop-like behavior is widely observed in the (magnetic-field-induced) phase transition from the $bc$-spiral AFM ($P_c$) state to the $ab$-spiral ($P_a$) one, as discussed adove for the $y=0.405$ sample.
These results suggest also that the $P_a$ phase of $y=0.52$ is not almost due to the $ab$-spiral AFM order but mainly due to the $E$-type AFM one.
The ferroelectric polarization along the $a$-axis of $y=0.405$, 0.52, and 0.55 at 5 K as a function of the poling electric field is shown in Fig.~\ref{fig3} (b).
In $y=0.405$, $P_a$ is easily saturated in a low poling field (about 100~kV/m) as well as $P_c$ of $y=0.55$ (Fig.~\ref{fig2} (b) inset). On the other hand, the saturation poling field of $P_a$ of $y=0.52$ is about 700~kV/m. The increase of saturation poling field is due to the increase of the $E$-type AFM nuclei embedded in the spiral AFM phase by increasing Y concentration $y$.
It is reasonable that a small amount of the $ab$-spiral AFM phase remains in the $E$-type AFM one at low temperatures, because $y=0.52$ is located near the phase boundary (see Fig.~\ref{fig4}).

We have summarized the obtained data into a magnetoelectric phase diagram for Eu$_{1-y}$Y$_y$MnO$_3$ crystals with $0 \le y \le 0.6$ (Fig.~\ref{fig4}). 
The ferroelectric transition temperature is steeply reducing with increasing $y$. 
In $y > 0.5$, the $P_c$ region drastically expands compared with that of $y=0.405$. 
As already discussed, in the $y=0.405$ sample,\cite{EuY_Noda} $P_c$ emerges below 25~K and disappears at 23~K\@, below which $P_a$ is observed. That is, the ferroelectric polarization flops from the $c$-axis to the $a$-axis at 23~K\@. 
This polarization flop is assigned to the orthogonal flop of the spin spiral plane from the $bc$ ($P_c$) to the $ab$ ($P_a$).
It should be noted that $P_c$ and $P_a$ are coexistent at low temperatures in $0.52\leq y\leq 0.6$ (right hatched area in Fig.~\ref{fig4}). 
The $P_a$ phase in the coexisting state should be attributed not to the $ab$-spiral AFM spin order but to the $E$-type AFM one.  
Because $P_c$ is easily saturated at a poling field of $\sim$ 100 kV/m, it is caused by the $bc$-spiral AFM order. 
On the other hand, $P_a$ is probably caused by the collinear $E$-type AFM order, because the poling electric field dependence of $P_a$ in $0.52\leq y\leq 0.6$ as shown in Fig.~\ref{fig2} (b), is very similar to that of $P_a$ due to the $E$-type AFM order in a polycrystalline form.\cite{Etype_poly} 
However, the presence of the $ab$-spiral AFM ($P_a$) phase cannot be ruled out as a possible source of $P_a$.
To confirm the appearance of the $E$-type AFM phase more directly, synchrotron X-ray and/or neutron diffraction measurements are required.

\section{Conclusion}
We have investigated the magnetic and dielectric properties of orthorhombic Eu$_{1-y}$Y$_y$MnO$_3$ ($0\leq y\leq 0.6$) single crystals without the presence of the 4$f$ magnetic moments of the rear-earth ions. 
In $y\geq 0.52$, two ferroelectric phases ($P\parallel a$ and $P\parallel c$) are coexistence at low temperatures. 
$P_c$ is induced by the noncollinear transverse spiral spin structure. 
On the other hand, $P_a$ is likely induced by the collinear $E$-type AFM order through the symmetric exchange striction term.
The two-phase coexisting state ($P_a+P_c$) would provide a novel route toward the control of dielectric (magnetic) properties by magnetic (electric) fields.

\section*{Acknowledgment}
This work was partly supported by Grant-in-Aid for JSPS Fellows from Japan Society for Promotion of Science.


\newpage

\begin{thebibliography}{99}
\bibitem{P101} From the ratio of the $P_{101}$ to the $(P_a+P_c)$ values measured in the series of $R$MnO$_3$, the correction coefficient to convert the observed $P_{101}$ value into the expected $(P_a+P_c)$ value was estimated to be $\frac{4\sqrt{2}}{3}$, which is roughly equal to the simple expectation of $\sqrt{2}$.
\end{thebibliography}
\end{document}